\documentclass[twocolumn, a4paper, 11pt,book]{archive}

\usepackage{version}
\usepackage{graphicx}

\usepackage[]{natbib}

\usepackage{txfonts}
\usepackage{booktabs}
\usepackage{multirow}

\def\lsim{\mathrel{\rlap{\lower4pt\hbox{\hskip1pt$\sim$}}
    \raise1pt\hbox{$<$}}}                
\def\gsim{\mathrel{\rlap{\lower4pt\hbox{\hskip1pt$\sim$}}
    \raise1pt\hbox{$>$}}}                
\usepackage{xcolor}
\usepackage{draftwatermark}
\SetWatermarkText{DRAFT}
\SetWatermarkScale{0.75}

\begin{document}

\twocolumn[{%
 \centering
%
  {\center \bf \huge Cosmic ray sputtering yield of interstellar H$_2$O ice mantles}\\
\vspace*{0.25cm}
{\Large Ice mantle thickness dependence}\\
\vspace*{0.25cm}

{\large   E. Dartois 
   \inst{1},
   \footnote{Part of the equipment used in this work has been financed by the French INSU-CNRS programme
Physique et Chimie du Milieu Interstellaire (PCMI).}
          M. Chabot \inst{2},
         T. Id Barkach \inst{2},
         H. Rothard \inst{3},
         B. Aug\'e \inst{4},
         A.N. Agnihotri  \inst{3},
        A. Domaracka \inst{3},
         P. Boduch \inst{3}\vspace*{0.5cm}
}\\
$^1$    Institut des Sciences Mol\'eculaires d'Orsay (ISMO), UMR8214, CNRS - Universit\'e de Paris-Sud, Universit\'e Paris-Saclay, 
B\^at 520, Rue Andr\'e Rivi\`ere, F-91405 Orsay, France\\
              \footnote{emmanuel.dartois@u-psud.fr}
 $^2$            Institut de Physique Nucl\'eaire d'Orsay (IPNO), CNRS-IN2P3, Universit\'e de Paris-Sud, Universit\'e Paris-Saclay, F-91406 Orsay, France
 $^3$            Centre de Recherche sur les Ions, les Mat\'eriaux et la Photonique, CIMAP-CIRIL-GANIL, Normandie Universit\'e, ENSICAEN, UNICAEN, CEA, CNRS, F-14000 Caen, France
 $^4$            Centre de Sciences Nucl\'eaires et de Sciences de la Mati\`ere (CSNSM), CNRS/IN2P3, Universit\'e de Paris-Sud, Universit\'e Paris-Saclay, F-91405 Orsay, France\\
\vspace*{0.5cm}
{\it \large To appear in Astronomy \& Astrophysics}
%
%
 \vspace*{0.5cm} 
 }]

\section*{Abstract}
   {Interstellar grain mantles present in dense interstellar clouds are in constant exchange with the gas phase via accretion and desorption mechanisms such as UV, X-ray photodesorption, cosmic ray induced sputtering, grain thermal fluctuations, and chemical reaction energy release. The relative importance of the various desorption mechanisms is of uttermost importance for astrophysical models to constrain the chemical evolution in such high density dense cloud regions.}
   {The sputtering yields for swift ions simulating the effects of cosmic rays are most often measured in the semi-infinite limit using thick ice targets with the determination of the effective yield per incident ion. In this experimental work we investigated the sputtering yield as a function of ice mantle thickness, 
exposed to Xe ions at 95MeV. The ion induced ice phase transformation and the sputtering yield were simultaneously monitored  by infrared spectroscopy and mass spectrometry.}
   {The sputtering yield is constant above a characteristic ice layer thickness and then starts to decrease below this thickness.
    An estimate of the typical sputtering depth corresponding to this length can be evaluated by comparing the infinite thickness yield to the column density where the onset of the sputtering yield decrease occurs. In these experiments the measured characteristic desorption depth corresponds to $\approx$ 30 ice layers. Assuming an effective cylindrical shape for the volume of sputtered molecules, the aspect ratio  is close to unity; in the semi-infinite ice film case this ratio is the diameter to height
of the cylinder. This result shows that most ejected molecules arise from a rather compact volume.
    The measured infinite thickness sputtering yield for water ice mantles scales as the square of the ion electronic stopping power (S$_e$, deposited energy per unit path length). Considering the experiments on insulators, we expect that the desorption depth dependence varies with S$_e^\alpha$, where $\alpha$$\sim$1. 
   Astrophysical models should take into account the thickness dependence constraints of these ice mantles in the interface regions when ices are close to their extinction threshold. In the very dense cloud regions, most of the water ice mantles are above this limit for the bulk of the cosmic rays.}


%

\section{Introduction}

Interstellar dust grains immersed in dense clouds acquire an ice mantle whose composition has been monitored by infrared spectroscopy for decades \citep[e.g.][]{Eiroa1983, Brooke1996, Brooke1999, Gibb2004, Dartois2005, Boogert2008, Oberg2011}.
When the number density is high enough, the accretion for very cold dust grains at about 10~K is efficient, and the condensation timescales become critical. Following various astrochemical models this timescale is short compared to the typical few tens of million of years lifetime of molecular clouds \citep[e.g.][]{Iglesias1977, Ruaud2016, Hincelin2016, Shingledecker2018}. Therefore most molecules should be trapped in the solid phase, whereas plenty of gas phase molecules and radicals are still observed.
For many years, efforts have been conducted in the laboratory to understand, constrain, and quantify the various processes counterbalancing the dust grain gas phase accretion, including UV and X-ray photodesorption \citep{Westley1995, Oberg2009, Munozcaro2010, Oberg2011, Fayolle2011, Fayolle2013, Munozcaro2016, Cruz-Diaz2016, Cruz-Diaz2018, Fillion2014, Dupuy2017,  Dupuy2018, Bertin2012, Bertin2016}. Some models also include progressively the transient nature of dust grain temperatures for mechanism including H$_2$ formation \citep[e.g.][]{Bron2014}. Cosmic rays also play a role among these desorption mechanisms.
The sputtering in the electronic regime of the stopping power, hereafter called electronic sputtering, by high energy heavy ions interacting with ice mantles has been the subject of numerous experimental studies \citep[e.g.][and references therein]{Seperuelo2009, Dartois2013, Boduch2015, Dartois2015, Mejia2015, Rothard2017} to provide input into astrophysical models. 
Experimentally, the use of heavy ions beams allow the measurement of the ice sputtering yields at high stopping power, complementing the numerous previous measurements at lower stopping power (e.g. Brown et al. 1984; Schou1987; Baragiola 2003; Fama et al. 2008, see also the book chapter \cite{Baragiola2013} and references therein). In addition, in this work the high sputtering yield of heavy ions is retrieved by monitoring (in situ) the evolution of the ice mantle thickness using infrared spectroscopy, which constrains the determination of electronic sputtering.
Once the dependence of the sputtering yield on stopping power is known over a large range, it can be implemented in a model including the elemental cosmic rays abundance and their corresponding differential flux at a given energy, as already discussed in for example \cite{Chabot2016} and \cite{Dartois2015}, and references therein.
It is mandatory to use the full range of experimentally determined stopping power dependence for a process to extrapolate to astrophysical media. 

In the experiments performed in this work, we use the general first order assumption of scaling factors for the stopping power interaction of particles with matter. 
It is remarkable to note that, for water ice, the quadratic dependence of electronic sputtering yield on electronic stopping power holds in a wide range. The range extends from low energy up to the stopping power of the heavy ion used in this study: that is at the onset of the electronic stopping power dominant contribution at tens/hundreds of keV/micron, with a yield of a few particle per incident ion (see references above),   to keV/nm and tens of thousands H$_2$O molecules sputtered per incident ion.

Cosmic ray elemental abundance is dominated by light ions, but heavy ions of intermediate masses (such as carbon and oxygen) up to higher masses (such as iron) contribute significantly to the processes if cross sections and/or yields rise quickly with the stopping power. As mentioned above, the electronic sputtering yield for ices varies quadratically with the stopping power. Furthermore, the stopping power S as a function of energy E for a particle of mass M and charge z evolves with respect to a proton of mass Mp as $\rm S(E,M,ze) = z^2 \times S(EMp/M,Mp,z=1)$. The stopping power is thus roughly quadratic with the atomic number. We also note that, owing to the cosmic ray acceleration processes, the heavy ions (Z$\geq$6) abundance with respect to protons is much larger than the elemental cosmic abundance \citep[by about an order of magnitude; see][]{George2009}. 
Therefore, in addition to the more abundant contribution of H$^+$ and He$^{2+}$ ions, heavy ions have to be included in astrophysical modelling. Experiments with swift heavy ions are, therefore, interesting technically to determine stopping power dependenciessuch as cross-sections, yields, and scaling factors. 
Also, rapid evolution of cross sections and/or yield with the stopping power can directly impact the astrophysical models with heavy ions own contribution. In the case of ice sputtering, the relative contribution of the various cosmic ray species then evolves in Z$^4$.
The energy dependence of the ice sputtering yield has been most of the time measured in the semi-infinite limit, i.e. often thick ice targets with respect to the number of sputtered molecules, for many ices.
Such experiments determine the total yield, but lack information on the surface to depth ratio of the volume probed, corresponding to the ejected molecules. The yield is dependent on the mantle thickness, as already observed in some experiments \cite[e.g.][CO$_2$ ice]{Mejia2015}, and \citep[e.g.][rare gas ices]{Schou1987}. 
The objective of the present study is to determine the thickness dependence of the yield for an amorphous compact water ice film irradiated with swift heavy ions to retrieve information on the depth/number of layers where the ejected molecules come from and sample the transition for the yield from bulk to thin film behaviour.
%
%
\begin{table*}
\caption{Nonlinear model results}             
\label{table:1}                
\begin{center}
\begin{tabular}{c c c c c c c l }     
\hline\hline       
\#ref           &Tmeas. & Tdep. &N$_0$$^a$                      &Thickness$^b$         &N$\rm_d$                       &Depth d(l$_d$)$^b$             &Y$_s^\infty$                 \\
                &K              & K             &[10$^{16}$cm$^{-2}$]   &\AA                         &[10$^{16}$cm$^{-2}$]   &\AA/layers                     &($\rm\times10^3$)      \\
\hline                    
27W1    & 10            & 10            &17.1$\pm$0.3                   &540$\pm$10             & 2.64$\pm$0.96           &84$\pm$30 (27$\pm$10)  &\it{20$\pm$2}          \\
27W2    & 10            & 100   &1.8$\pm$0.1                    &58$\pm$3               & > 0.23                          & n.a.                          & > 7.5                         \\
27W3    & 10            & 10            &2.9$\pm$0.1                    &92$\pm$3               & >1.93                           & n.a.                          &>16                    \\
29W1    & 10            & 100   &60.1$\pm$1                     &1924$\pm$34            & n.a.                            & n.a.                          &21$\pm$2                 \\
30W2    & 10            & 10            &59.7$\pm$1                     &1880$\pm$32            &3.48$\pm$0.37          &111$\pm$13 (35$\pm$4)      &22$\pm$1.2             \\
%
%
\hline                  
\end{tabular}
\end{center}
Tmeas. and Tdep. are the ice temperature during measurement and deposition, respectively. N$_0$ and thickness are the t=0 initial ice column density and film thickness. Depth d(l$_d$) is the sputtering depth (depth in number of ice monolayers) and N$\rm_d$ the corresponding ice column density. Y$_s^\infty$ is the electronic sputtering yield for a semi-infinite ice film.
Uncertainties given are statistical uncertainties calculated on the model fit.
$^a$  From the fit and assuming an integrated absorption cross section of $\rm A=2.2\times 10^{-16}$ cm/molecule \citep[][]{Leto2003, Mejia2015b} for the amorphous compact ice. 
$^b$ The density considered for the ice thickness and number of molecular layers determination is 0.93~g/cm$^3$ \citep[][]{Feistel2006}
\end{table*}
%
%
%
%
\begin{figure}[htbp]
\begin{center}
\includegraphics[width=\columnwidth,angle=0]{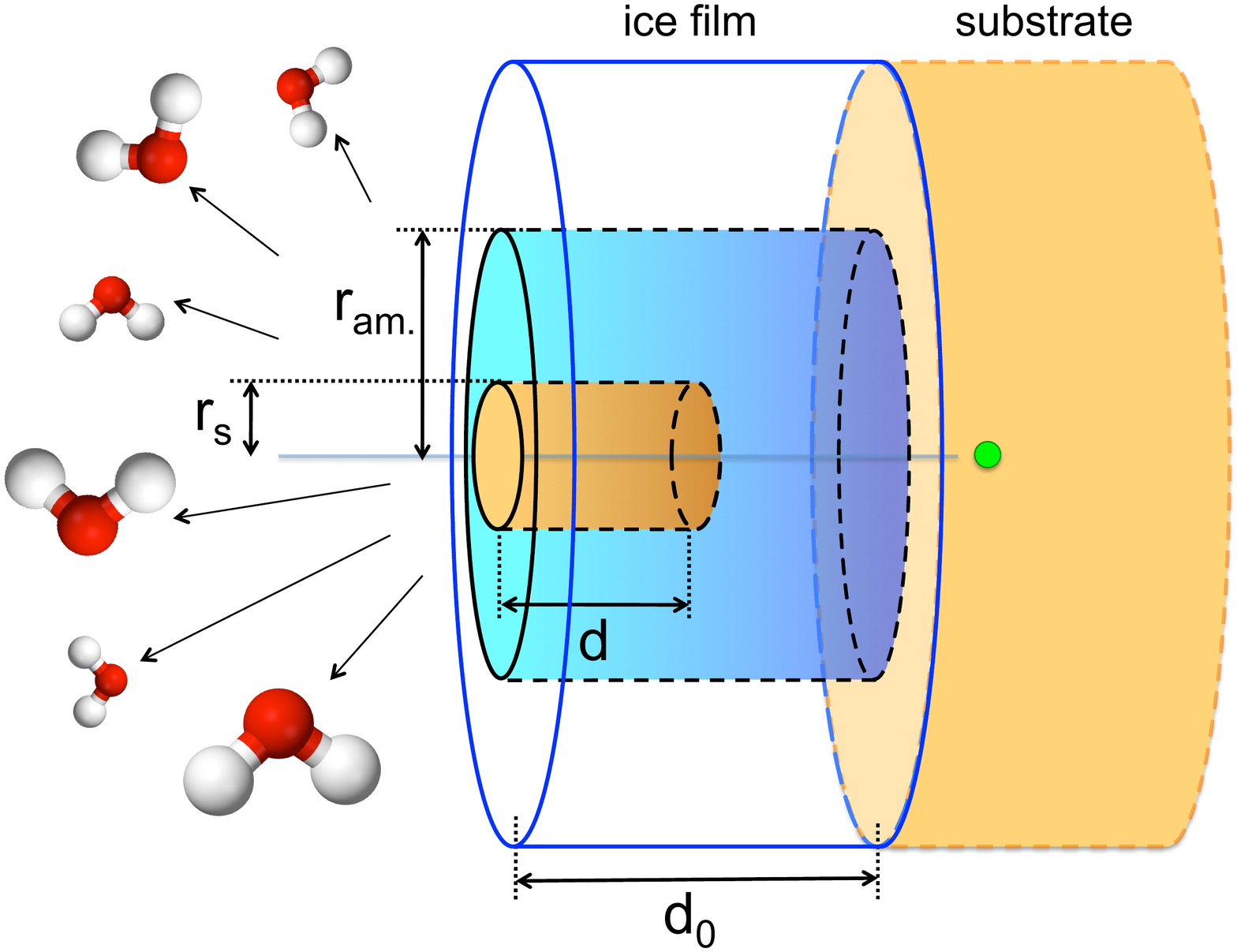}
\caption{Schematic view of the sputtering yield experimental and effective cylinder geometry model. d$\rm_0$ (N$\rm_0$) corresponds to the initial ice film thickness (column density) and d (N$\rm_d$) to the sputtering depth (column density) probed by an individual ion incident. $\rm r_s$ correspond to the effective sputtering cylinder radius, and $\rm r_{am.}$ to the amorphisation radius.}
\label{schema}
\end{center}
\end{figure}
\section{Experiments}
Swift ion irradiation experiments were performed at the heavy-ion accelerator Grand Acc\'el\'erateur National d'Ions Lourds (GANIL; Caen, France).
Heavy ion projectiles were delivered on the IRRSUD beam line\footnote{http://pro.ganil-spiral2.eu/laboratory/experimental-areas}. The IGLIAS (Irradiation de GLaces d'Int\'er\^et AStrophysique) facility, a vacuum chamber (10$^{-9}$ mbar under our experimental conditions) holding an infrared transmitting substrate that can be cryocooled down to about 8 K, is coupled to the beam line. Ice films are produced by placing a cold window substrate in front of a deposition line. Amorphous water ice films were condensed at 10~K (or 100~K) on the window, from the vapour phase, and kept at 10~K during the irradiations.
Some experiments were conducted with an ice mantle grown at 10~K, i.e. starting with an amorphous ice. A few experiments were performed with an ice mantle grown at 100~K, and irradiated at 10~K. The film deposited at 100~K has a non-porous structure \cite[e.g.][]{Stevenson1999} closer to the compact-amorphous state resulting from ice compaction induced by ion irradiation in the beginning of irradiation, for comparison.
Details of the experimental setup are given in \cite{Auge2018}.
The ion flux, set between 10$^7$ to 10$^9$ ions/cm$^2$/s is monitored on-line using the current measured on the beam entrance slits defining the aperture. 
The relation between the current at various slits apertures and the flux is calibrated before the experiments using a Faraday cup inserted in front of the sample chamber. 
The thin ice films deposited allow the ion beam to pass through the film with an almost constant energy loss per unit path length.
A Bruker FTIR spectrometer (Vertex 70v) with a spectral resolution of 1 cm$^{-1}$ was used to monitor infrared film transmittance. The evolution of the infrared spectra was recorded as a function of the ion fluence.
Irradiation is performed at normal incidence, whereas the infrared transmittance spectra are recorded simultaneously at 12$\rm^o$ of incidence.
A sweeping device allows for uniform and homogeneous ion irradiation over the target surface.
The flux delivered on IRRSUD was stable during the experiments except for a gradual shift for the first experiment. 
The sweeping rate was not optimised for this particular experiment and the overall sample irradiation probed by the IR spectrometer received about 40\% less ions than expected. For this first experiment the fluence was rescaled using the infrared spectra evolution slope observed for the other experiments during the thick film phase.
The projectiles ($\rm ^{136}Xe^{23+}$ at 0.7 MeV/u) electronic stopping power is close to 8keV/nm, as calculated from the SRIM package \citep{Ziegler2010}, adopting an ice density of 0.93 g/cm$^3$. The corresponding projectile range is of the order of 22 $\mu$m, which is well above the ice film thicknesses used in this study.
This stopping power value lies in a range close to previous thick film experiments \citep{Dartois2015}. This previously determined yield dependence with stopping power predicts a sputtering yield of 22700 sputtered H$_2$O/ion (although with a large error bar).
Mass measurements were performed simultaneously using a microvision2 mks quadrupole mass spectrometer (QMS) recording masses from 1 to 100 amu. In some experiments only the water fragmentation pattern in the 16 to 18 amu range was recorded to increase the sampling rate. The beam was regularly stopped during a few cycles in order to record and follow the evolution of the QMS and chamber background signal.
The QMS spectra presented are background subtracted.
A summary of the ice film parameters such as thickness, deposition, and irradiation temperatures is given in Table~\ref{table:1}.
\section{Model}
\label{model}
As discussed previously \citep{Dartois2015}, the column density evolution of the water ice molecules submitted to ion irradiation can be described, to first order, as a function of ion fluence (F) by a coupled set of differential equations as follows:
\begin{equation} 
\left\lbrace
\begin{array}{ccl}
\rm dN/dF       &\rm =  &\rm -\sigma_d N - Y_s \times f\\
\rm dw_a/dF     &\rm =          &\rm -\sigma_{C} \times w_a
\end{array}\right.
.\end{equation}
The $\rm -\sigma_d N$ and $\rm Y_s \times f$ terms in the top part of eq.1 are related to the destruction and sputtering contributions to the evolution of the ice column density. The value 
N is the total H$_2$O column density, $\rm \sigma_d$ the ice effective radiolytic destruction cross-section (cm$^{2}$), and $\rm Y_s$ the H$_2$O sputtering yield (H$_2$O/ion), multiplied, to first order, by the relative concentration f of water ice molecules with respect to the total number of molecules/radicals in the ice film. In the second differential equation, $\rm \sigma_{C}$ is the ice compaction cross section (amorphous to compact-amorphous ice), and $\rm w_a$ the amorphous ice fraction.\\
Pure water ice is very radiolytically resistant as compared to other ice mantle ices and/or mixtures. One of the main products formed by irradiation is H$_2$O$_2$, which saturates to a few percent of water-ice molecules, by number, at doses above $\sim$10 eV/molecule (see e.g. \citealt{Moore2000} for the production yield of H$_2$O$_2$). The same low efficiency is observed for far-ultraviolet (FUV) irradiation of pure ice by \cite{Gerakines1996}, among others.
Considering the radiolytic resilience of pure water ice, we impose that f remains close to unity. For our considered experimental fluences and ice film thicknesses, if radiolysis is negligible, then the evolution of the water ice spectra is dominated by  amorphous to compact-amorphous phase change and sputtering. 

When the water ice film is thin (column density $\rm N\lesssim N_D$) the removal of water molecules by sputtering follows a direct impact model, i.e. all the molecules within the sputtering area defined by a sputtering effective cylinder (see Fig.\ref{schema}) are removed from the surface. 
The apparent sputtering yield, as a function of thickness, can therefore be modelled to first order by
\begin{equation} 
\begin{array}{ccl}
\rm Y_s(N)              &\rm =  &\rm Y_s^{\infty} \left( 1 - e^{-\frac{N}{N_D}} \right)\\
\end{array}
\label{equation_yield}
,\end{equation}
where $\rm Y_s^{\infty}$ is the semi-infinite ice film sputtering yield, and $\rm N_D$ is the ice column density corresponding to the maximum depth probed by the sputtering by an individual ion incident.
The coupled set of equation to solve for an infrared analysis is therefore
\begin{equation} 
\left\lbrace
\begin{array}{ccl}
\rm dN/dF               &\rm =  &\rm -Y_s^{\infty} \left( 1 - e^{-\frac{N}{N_D}} \right)\\
\rm dw_a/dF     &\rm =          &\rm -\sigma_{C} \times w_a
\end{array}\right.
.\end{equation}
The solutions to these equations come down to
\begin{equation} 
\left\lbrace
\begin{array}{ccl}
\rm N(F)                &\rm =  &\rm N_D \times ln \left( e^{-\frac{Y_s^{\infty} F}{N_D}+ln\left(e^{\frac{N_0}{N_D}}-1\right)}+1 \right)\\
\rm w_a                 &\rm =          &\rm exp(-\sigma_{C} \times F) 
\end{array}\right.
,\end{equation}
where $\rm w_c=(1-w_a)$ is the amorphous compacted ice fraction.
The column densities of the molecules are followed experimentally in the infrared via the integral of the optical depth ($\rm \tau$) of a vibrational mode, taken over the band frequency range. The band strength value (A, in cm/molecule) for a vibrational mode and ice structure (compact-amorphous or amorphous) has to be considered.  During the irradiation, the phase change from an amorphous to a compact-amorphous state induces a variation in the band strengths \citep[see e.g. ][]{Leto2003}. Therefore, this has to be compensated in order to extract the true column density, i.e.
\begin{equation} 
\rm N = \frac{1}{A_c}\int^{\bar{\nu}2}_{\bar{\nu}1} \tau(\bar{\nu})d\bar{\nu} \; [w_c+ \alpha(1-w_c)] ; \alpha=A_c/A_{am} ,
\end{equation}
where $\rm A_{am}$ and $\rm A_c$ are the band strength values for amorphous and compact-amorphous ice (integrated over the same wavenumber range).

With the particular projectile ion considered, the amorphous to compact ice phase change occurs with a cross section of a few 10$^{-11}$ cm$^2$ \citep{Dartois2013} and thus for the thicker ice films considered, after a fluence of a few 10$^{-11}$ cm$^{-2}$ the measured integrated absorbance variations directly reflect the ice thickness evolution.
%
\begin{figure}[htbp]
\begin{center}
\includegraphics[width=\columnwidth,angle=0]{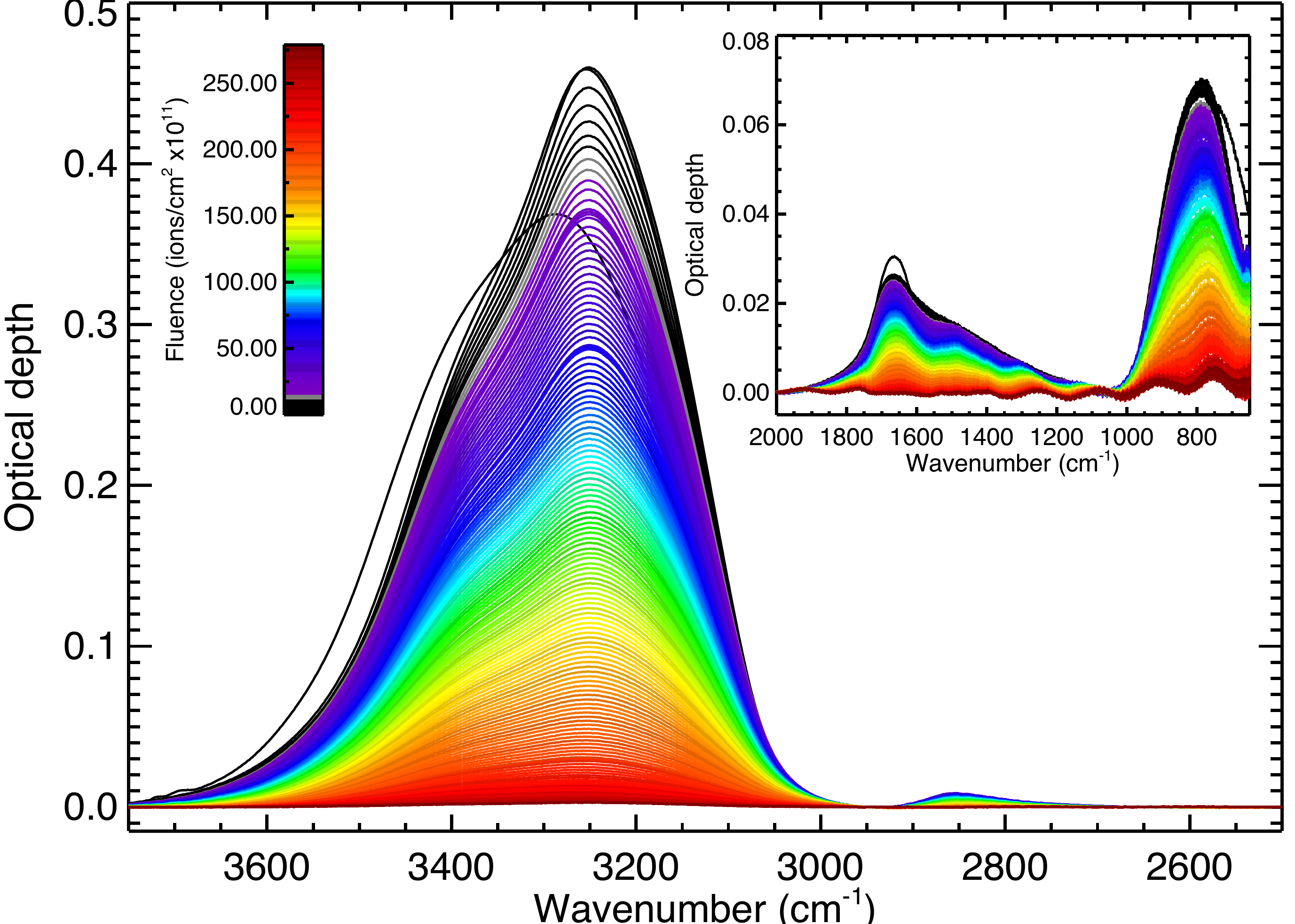}
\caption{Infrared optical depth spectral evolution as a function of Xe ions fluence during the 30W2 experiment. The main absorption corresponds to the water ice OH stretch and the insert shows the bending and libration modes.}
\label{30W2_IR}
\end{center}
\end{figure}
%
\begin{figure*}[htbp]
\begin{center}
\includegraphics[width=1.9\columnwidth,angle=0]{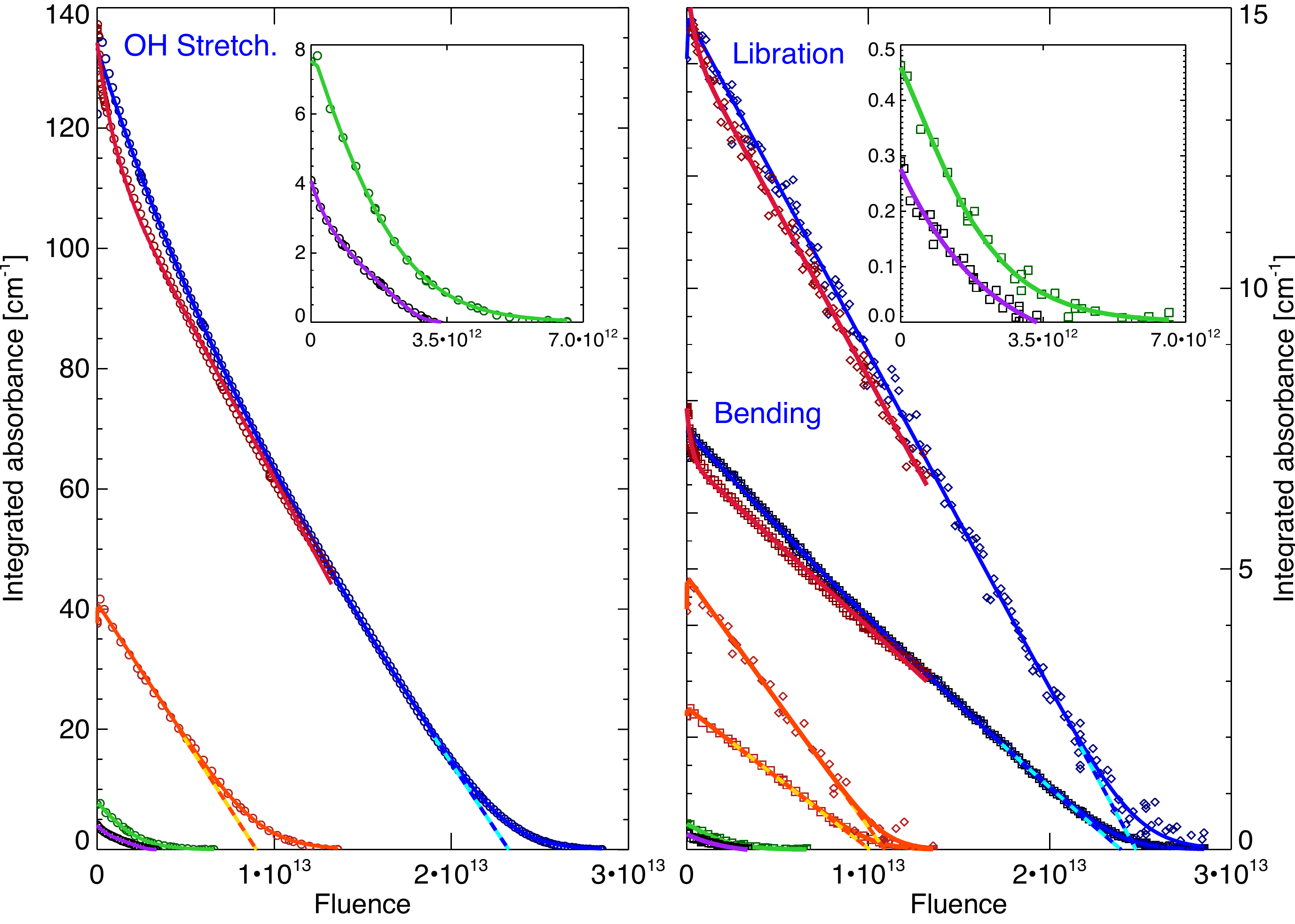}
\caption{Infrared integrated absorbance measured as a function of ion fluence for ice films of various initial thicknesses, measured at 10~K in the OH stretching mode (circles, left), bending and libration modes (right, squares and diamond, respectively). The different colours correspond to the experiments summarised in Table~\ref{table:1}: orange (27W1), purple (27W2), green (27W3), red (29W1), and blue (30W2). The model fits are superimposed with the same colour for the various IR modes corresponding to a single experiment. The dashed lines represent models in which the sputtering yield would be independent of the thickness.}
\label{IR_general}
\end{center}
\end{figure*}
%
%
\section{Results and discussion}
\label{results}
\begin{figure}[htbp]
\begin{center}
\includegraphics[height=\columnwidth,angle=90]{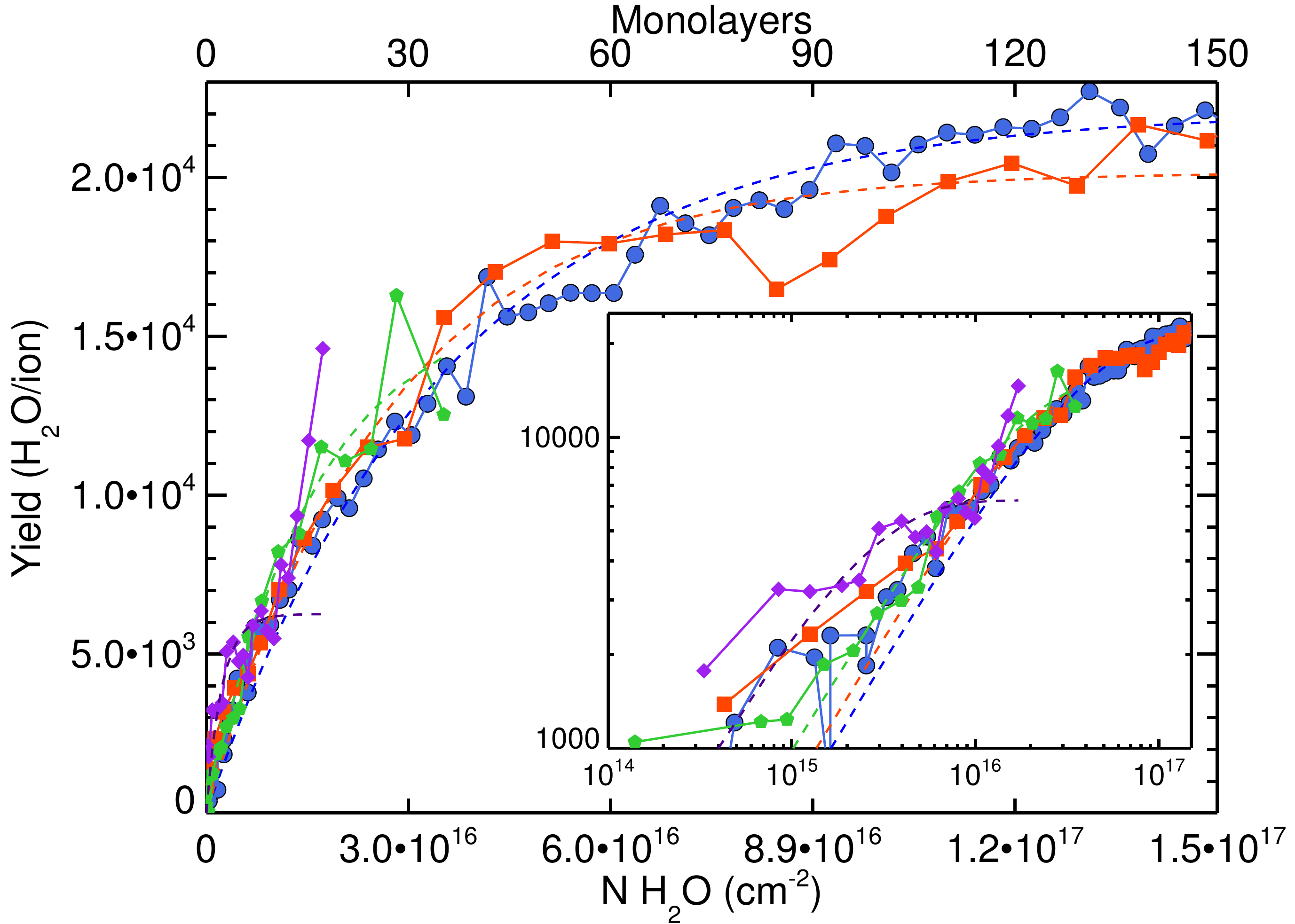}
\caption{Sputtering yield dependence calculated from the infrared spectral analysis, as a function of ice film column density. Dotted line correspond to a fit using equation~\ref{equation_yield}. The colour coding is the same as for Fig. \ref{IR_general}. The upper scale gives the corresponding number of ice monolayers assuming the ice density given in Table \ref{table:1}.}
\label{yields}
\end{center}
\end{figure}

The baseline corrected optical depth infrared spectra for the 30W2 experiment, as a function of ion fluence, are shown in Fig.\ref{30W2_IR}. 
The infrared integrated absorbance of the OH stretching, bending and libration modes extracted from such measurements for all the experiments, corresponding to ice films of different initial thickness, are reported as a function of ion fluence in Fig.\ref{IR_general}. The signal-to-noise ratio for the libration mode integral suffers from higher variations due to uncertainties in the baseline determination at low frequencies.
The colours correspond to the experiments reported in Table~\ref{table:1}. Model fits are superimposed with the same colour for the various infrared modes corresponding to a single experiment. At the very beginning of the irradiation of thick films, the integral evolves due to phase change and then lies on a straight line corresponding to the semi-infinite thick film sputtering. When the films get thinner, the signal evolves with an exponential behaviour as expected from equation~\ref{equation_yield}. The dashed lines represent models in which the sputtering yield would remain independent of the thickness, to delineate the integrated absorbance when the deviation from semi-infinite behaviour starts. For two of the films, the irradiation already starts in the thin film limit and we cannot extract the semi-infinite asymptotic behaviour.
%
\begin{figure}[htbp]
\begin{center}
\includegraphics[width=\columnwidth,angle=0]{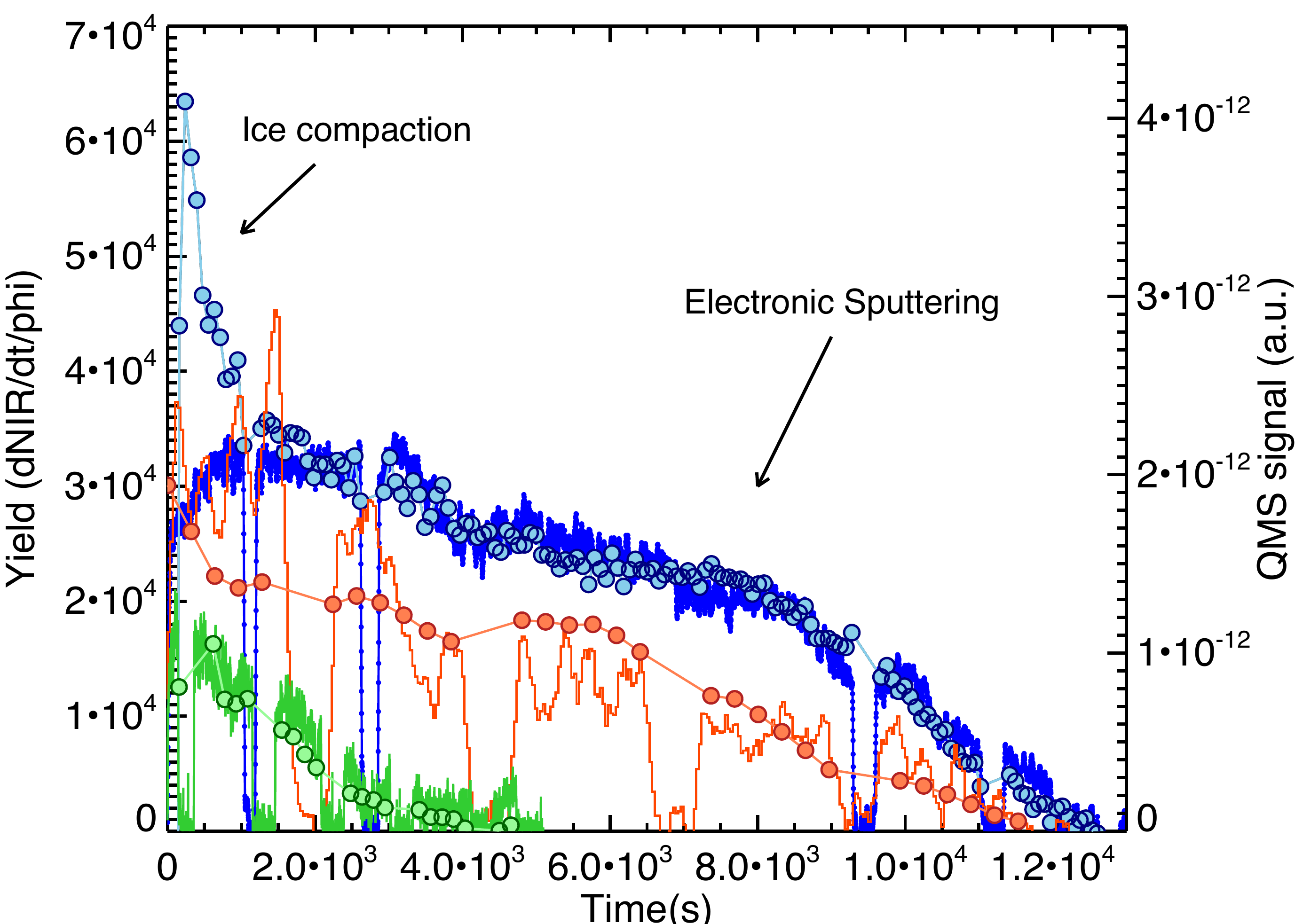}
\caption{Temporal evolution of the signal recorded with the QMS at mass 18 for three of the experiments (lines), scaled and compared to the derivative of the IR signal with respect to the ion flux (circles). The  colours correspond to the experiments summarised in Table~\ref{table:1}: orange (27W1), green (27W3), blue (30W2). The breaks correspond to ion beam stops in order to monitor properly the mass spectrometer background signal evolution.}
\label{QMS}
\end{center}
\end{figure}
%
Mass spectrometer measurements were performed simultaneously. The temporal evolution of the signal recorded with the QMS at mass 18 (H$_2$O$^+$) for three of the experiments
are shown in Fig.~\ref{QMS}, and scaled to be compared to the disappearance of water ice molecules recorded in the infrared (using the integral of the OH stretching mode) per incident ion, corresponding to the sputtering yield. A good correlation between the two signals can be observed, except in the beginning of the irradiation for the infrared. This is easily explained by the phase transformation from the initial ice phase to compact-amorphous ice, inducing variations of the OH stretching mode oscillator strength. We note that the sputtering yield monitored by the QMS during this phase remains rather constant and thus the sputtering yield does not vary as much as the phase transformation observed in the infrared would suggest.
From the model fits, the semi-infinite sputtering yield and the column density $\rm N_D$ at which the thin film limit is reached, can both be extracted. Assuming an ice density of 0.93~g/cm$^3$ \citep[][]{Feistel2006}, the $\rm N_D$ value translates into a sputtering depth d, and a number of ice layers $\rm l_d$. These figures are reported in Table~\ref{table:1}. The yield measured as a function of the film thickness is reported in Fig.~\ref{yields}.

The radial symmetry of the energy transferred in the track by the incident ion motivates the choice of the effective cylinder geometry adopted (shown in Fig.~\ref{schema}) to explore the stopping power dependence of the yield.
To figure out how the characteristic lengths transverse to the beam and along the track evolve for the sputtered volume, we define the aspect ratio (A.R.) as the ratio of depth (d) to the diameter ($\rm 2r_s$) assuming the effective cylindrical shape for the track and sputtered volume, i.e. 
A.R.$\rm = d / (2r_s )= d / (2 \sqrt{\rm V/\pi d}) = l_d / (2 \sqrt{\rm Y_s^\infty/\pi*l_d}) = l_d^{3/2} / (2 \sqrt{\rm Y_s^\infty/\pi})$. 
From Table~\ref{table:1}, we calculate A.R.$\approx$0.88 for the 27W1 and $\approx$1.23 for the 30W2 experiments, respectively, implying that the sputtered volume is not elongated along the ion track.

In a recent experiment performed on the sputtering of CO$_2$ ice by $\rm^{132}Xe^{21+}$ ions at 630~MeV, \cite{Mejia2015} measured a reduced sputtering yield for a thin film as compared to the semi-infinite thick ice measurements. The stopping power of this ion is similar to that used in our study for water ice, the semi-infinite CO$_2$ sputtering yield is about an order of magnitude higher. This study did not explore the thickness dependence of the yield, but the comparison between thin and thick film yields seems to show that the aspect ratio is higher than for H$_2$O, and further experiments will be conducted to explore the behaviour for this apolar ice mantle constituent.

%

Experiments and thermal spike models of the ion track induced phase transformation in insulators predict a dependence of the radius r of the cross section evolving as $\rm r \sim \sqrt{S_e}$, where $\rm S_e=dE/dx$ is the deposited energy per unit path length \citep[e.g.][]{Lang2015,Toulemonde2000,Szenes1997}, and with a threshold in $\rm S_e^{th}$ to be determined. 
At the same time, the measured semi-infinite (thick film) sputtering yield for water ice mantles (i.e. corresponding to the total volume) scales as the power of two of the ion electronic stopping power \citep[$\rm Y_s^{\infty}\propto S_e^{2}$;][]{Dartois2015,Mejia2015}. 
Combining these dependences, in the electronic sputtering regime considered in these experiments, we therefore expect the sputtering depth to scale almost linearly with the stopping power, and the aspect ratio to scale with the square root of the stopping power.
Based on the figures given in Table~\ref{table:1}, if we extrapolate the linear dependency the monolayer desorption depth should lie close to a stopping power of about 250~keV/$\mu$m.

Previous measurements on the compaction of amorphous ice have led to the evaluation of the cross section of the process as a function of the stopping power \citep[][]{Dartois2013}. At $\rm S_e=8~keV/nm$, the compaction cross section is around $\rm 10^5\AA^2$ starting from an amorphous ice (ASW). An estimate of a sputtering cross section can be inferred from our measurements $\rm \sigma_s\approx V/d$, where V is the volume occupied by $\rm Y_s^{\infty}$ molecules. With the same density as adopted previously, we obtain $\rm \sigma_s\approx5.4-7.5\times10^4\AA^2$. The amorphisation radius $\rm r_{am.}$ would therefore be about 1.2 to 1.4 times larger than the sputtering radius $\rm r_s$.

In the context of astrophysical observations, the correlation between observed ice mantle optical depth, in the OH stretching mode of water ice, and silicates extinction allows us to draw an overall ice to silicates volume ratio of about unity \cite[e.g.][]{Dartois2006}. This value is the lower limit for the ice/core volume ratio as it implies that all the grains are coated with an ice mantle along the line of sight. Assuming that the grain size distribution is at least larger than the diffuse interstellar medium so-called MRN \citep[][]{Mathis1977} assumed power law; $\rm dn/da \propto a^{-3.5}$ with the grain radius from a few nm to submicron size; and n the number of grains, or more modern size distributions, if any eventual grain growth has begun;
the ice mantle thickness is therefore above $\sim$100~$\AA$ in well shielded dense cloud regions, i.e. a few A$_V$ after the ice mantles threshold. For the majority of cosmic ray energies, the stopping power corresponds to lower values and the yield is then dominated by the infinite thickness yield in such high A$_V$ lines  of sight for water ice mantles. 

\section{Conclusions}

We have measured the swift heavy ion compact-amorphous water ice mantle sputtering yield at 10~K and its dependence with the ice thickness.
These measurements allow us to constrain the sputtering depth probed by the incident ion.
In these experiments the measured characteristic desorption depth is of $\rm100\pm20\AA$, corresponding roughly to 30 ice layers. 
Assuming an effective cylindrical shape for the volume of sputtered molecules, the aspect ratio is close to unity;   in the semi-infinite
ice film case this ratio is the height to diameter of the cylinder. This shows that most of the ejected molecules arise from a rather compact volume.
The measured infinite thickness sputtering yield for water ice mantles scales as the square of the ion electronic stopping power (S$_e$, deposited energy per unit path length). Considering the experiments on insulators, we expect that the desorption depth dependence varies with S$_e^\alpha$, with $\alpha$$\sim$1. 
Astrophysical models should take into account these ice mantle thickness dependence constraints, in particular at the interface regions and onset of ice formation, i.e. when ice is close to the measured extinction threshold. 
The present measurements performed at a single electronic stopping power provide an anchor point to estimate the number of ice sputtered layers as a function of the energy. Additional measurements varying the deposited energy will be performed to constrain the energy dependence of the parameters. 
\begin{acknowledgements}
This work was supported by the Programme National "Physique et Chimie du Milieu Interstellaire" (PCMI) of CNRS/INSU with INC/INP co-funded by CEA and CNES. 
Experiments performed at GANIL. We thank T. Madi, T. Been, J.-M. Ramillon, F. Ropars, and P. Voivenel for their invaluable technical assistance. 
A.N. Agnihotri acknowledges funding from INSERM-INCA (Grant BIORAD) and R\'egion Normandie fonds Europ\'een de d\'eveloppement r\'egional-FEDER Programmation 2014-2020.
We would like to warmly thank the editor Mario Tafalla and an anonymous referee for the suggestions that have enhanced the content and improved this article, as well as the language editor for proofreading the article.
\end{acknowledgements}


\end{document}